\documentclass[journal,onecolumn,draftcls,letter,12pt]{IEEEtran}
\ifCLASSINFOpdf
\else
\fi
\usepackage[cmex10]{amsmath}
\usepackage{tikz,,cite, graphicx}
\begin{document}
\title{Resource Allocation for Selection-Based Cooperative OFDM Networks}
\author{\IEEEauthorblockN{Kianoush Hosseini and Raviraj Adve\\}
\IEEEauthorblockA{Department of Electrical and
Computer Engineering, University of Toronto\\  10 King's College Road,
Toronto, ON, Canada M5S 3G4}\\
 Email: \{kianoush, rsadve\}@comm.utoronto.ca}


\maketitle

\begin{abstract}
This paper considers resource allocation to achieve max-min fairness in
a selection-based orthogonal frequency division multiplexing network
wherein source nodes are assisted by fixed decode-and-forward relays. The joint problem of transmission strategy selection, relay assignment, and power allocation is a combinatorial problem with exponential complexity. To develop effective solutions to these questions, we
approach these problems in two stages. The first set of problems assume
ideal source-relay channels; this simplification helps illustrate our general
methodology and also why our solutions provide tight bounds. We then
formulate the general problem of transmission strategy selection, relay
assignment, and power allocation at the sources and relays considering
all communication channels, i.e., finite power source-relay channels.
In both sets of problems mentioned so far, transmissions over
subcarriers are assumed to be independent. However, given the attendant
problems of synchronization and the implementation using a FFT/IFFT
pair, resource allocation at the subcarrier level appears impractical.
We, therefore, consider resource allocation at the level of an entire
OFDM block. While optimal resource management requires an exhaustive
search, we develop tight bounds with lower complexity. Finally, we
propose a decentralized block-based relaying scheme. Simulation results
using the COST-231 channel model show that this scheme yields
close-to-optimal performance while offering many computational
benefits.
\end{abstract}

\begin{IEEEkeywords}
Cooperative communication, orthogonal frequency division multiplex
(OFDM), resource allocation.
\end{IEEEkeywords}

\IEEEpeerreviewmaketitle
\section{Introduction}
In a cooperative wireless network, geographically distributed nodes
share the available resources to achieve the benefits of multiple-input
multiple-output systems and combat the impact of fading via relaying.
The initial work in~\cite{SEA03,LTW04,LW03} sparked much research
activity in this area. Of specific interest here is the
decode-and-forward (DF) protocol where the relay node decodes and
re-encodes the source's data~\cite{LTW04}. If multiple relays are
available, selection, wherein sources choose one ``best" relay, has
been shown to provide almost all the benefits of the cooperative
diversity while minimizing overhead. Most importantly, selection avoids
issues of synchronization across relays; selection-based cooperation
has now been studied in various
contexts~\cite{BKRL06,BA08,ZAL07,MK08,CAE07}. However, relay selection
becomes more crucial in multi-source networks when simultaneous data
flows are allowed. Since each relay must split its available power
amongst all the source nodes it supports, the individually optimal
relay allocation scheme may not be globally optimal. Optimal relay
assignment is a combinatorial optimization problem with exponential
complexity. Without addressing power allocation at the relays,
in~\cite{BA08} the authors present low complexity relay selection
schemes for multi-source networks. In~\cite{KA09}, the authors extend
this to include power allocation in a single-carrier cellular network,
but assuming an ideal source-relay channel.

In a separate track, orthogonal frequency division multiplexing (OFDM/OFDMA)
is an increasingly popular technique to mitigate the impact of
multipath fading and enables high data rates for current and emerging
wireless communication technologies. Furthermore, because each
subcarrier experiences a different channel realization, resource
allocation can significantly enhance
performance~\cite{WCLM99,RC00,JB03,SAE05}. OFDM benefits from the
crucial implementation advantage that the transmitted signal can be
obtained from an Inverse Fast Fourier Transform (IFFT) of the data.
This IFFT is paired with a FFT at the receiver. However, as we will
see, this pairing also restricts how nodes can cooperate.

In the context of cooperative OFDM networks, optimal relay assignment and dynamic subcarrier
and power allocation have received significant attention.
In~\cite{LL06} Li et al.~developed a graph-theoretical approach to
maximize the sum rate under fairness constraints; here fairness is
imposed by limiting the number of sources a single relay can help. The
work in~\cite{LLWLL09} maximizes the minimum rate in a two-hop
cooperative network while allowing for subcarrier permutation.
In~\cite{NY07}, Ng and Yu constructed a utility maximization framework
to solve for relay selection, relaying strategies, and power allocation
in cellular OFDMA access-based
networks. Using the decomposition method and assuming a finite discrete
set of rates, they use an exhaustive search to solve the optimization
problem. The same approach is employed in~\cite{WM08} in order to minimize
the required power subject to data rate constraints on each flow. The
authors of~\cite{CLW09} proposed a resource allocation scheme for a
two-hop clustered-based cellular network with relays chosen \textit{a
priori}.

A key concern is that all these works deal with the OFDM transmission
on a per-subcarrier basis, i.e., as if each were an
\textit{independent} transmission. In OFDM-based networks, the raw data is encoded for error control before data
modulation and the IFFT; the data is therefore spread over all
subcarriers. In turn, this implies that DF requires decoding \emph{all
subcarriers}. Most of the subcarrier-based resource allocation is,
therefore, theoretically optimal, but impractical. Hence, relay selection and resource allocation must happen over the
entire OFDM block of subcarriers.

In this paper, we consider a selection-based cooperative OFDM mesh network
of access points (AP) where relays use the DF protocol. We begin with
the assumption that all relay nodes can always decode each individual
data streams, i.e., we assume that the relay can always decode the source transmissions without error. While this may be valid in a few practical scenarios,
e.g., when relays are installed close enough to the source nodes, this
is clearly not a universally valid assumption. However, as we will see,
the solution based on this simplifying assumption provides useful
insights into finding a near-optimal solution for a subcarrier-based
selection scheme. 

Building on the work in~\cite{KA09}, here we prove that subcarrier selection at the relays 
is the optimal assignment for \textit{most, not all,} subcarriers. This contradicts the results in~\cite{PNB08}
which claims that subcarrier based selection is the optimal strategy.
Furthermore, using the Karush-Kuhn-Tucker (KKT) conditions,
we characterize an upper bound to the original subcarrier-based problem
which leads to joint relay and power allocation for each subcarrier. We
also derive a simple tight lower bound on the solution of the original
problem. We then deal with selection for an entire OFDM block and
propose a simple selection scheme, but with performance close to using
an exhaustive search and not much different from the per-subcarrier
relaying scheme. To the best of our knowledge, there has been no
published work on selection and resource allocation at the level of an
entire OFDM block.

In the next section, we solve the general form of the relay selection
and resource allocation problem for OFDM-based networks irrespective of
the positions of the relays, i.e., unlike previous works, we take
source-to-relay (S-R), source-to-destination (S-D), and
relay-to-destination (R-D) channels into account. Furthermore, our
scheme allows for \emph{direct transmission (no cooperation)} if it
is optimal. By introducing time-sharing coefficients, we transform the original combinatorial problem into a standard convex optimization
problem resulting in an upper bound on performance. In addition, using
the same approach, we formulate block-level selection for multi-source
networks and develop an upper bound to the achievable rate. A
tight lower bound for subcarrier-based (respectively block-based)
schemes can be achieved by enforcing the selection constraint, i.e.,
each subcarrier (resp.~block) is transmitted either via a single
relay or directly to the destination. Finally, we propose a distributed, decentralized, 
selection scheme which offers large computational advantages, but with
close-to-optimal performance. We emphasize that unlike most previous
works, neither the transmission strategy nor the relay nodes are chosen
\textit{a priori}.

The remainder of this paper is organized as follows.
Section~\ref{sec:SysModel} presents our system model in some detail.
Section~\ref{sec:perfect} investigates node selection and resource
allocation under the assumption that all relays can decode.
Section~\ref{sec:imperfect} deals with the optimization problem for
both subcarrier-based and block-based schemes by considering all
communication channels while taking both selection and per-node
power constraints into account. Section~\ref{sec:Simulation} presents
simulation results that quantify the performance of different
relaying and resource allocation algorithms. Finally, we wrap up this
paper in Section~\ref{sec:conclusion} with some conclusions.

\section{System Model}\label{sec:SysModel}
\begin{figure}
\centering
\begin{tikzpicture}[scale=1.1]
 \tikzstyle{post}=[very thick, ->, >=stealth]
            \node (source2) at (0,0) [ball color=white,circle,fill=white,thick,
            minimum size = 1.5cm, text=black] {$\mathcal{K}_2$};
            \node (source1) at (0,2) [ball color=white,circle,fill=white,thick,
            minimum size = 1.5cm, text=black] {$\mathcal{K}_1$};
            \node (source3) at (0,-2) [ball color=white,circle,fill=white,thick,
            minimum size = 1.5cm, text=black] {$\mathcal{K}_3$};
            \node (dest1)   at (6,2) [ball color=white,circle,fill=white,thick,
            minimum size = 1.5cm, text=black] {$\mathcal{D}_1$};
            \node (dest2)   at (6,0) [ball color=white,circle,fill=white,thick,
            minimum size = 1.5cm, text=black] {$\mathcal{D}_2$};
            \node (dest3)   at (6,-2) [ball color=white,circle,fill=white,thick,
            minimum size = 1.5cm, text=black] {$\mathcal{D}_3$};
            \node (relay1)  at (3,1) [ball color=white,circle,fill=white,thick,
            minimum size = 1.5cm, text=black] {$\mathcal{R}_1$};
            \node (relay2)  at (3,-1) [ball color=white,circle,fill=white,thick,
            minimum size = 1.5cm, text=black] {$\mathcal{R}_2$};

            \path[post] (source1) edge [color = black,thick] (relay1)
                      (source1) edge [color = black] (relay2)
                      (source2) edge [color = black] (relay1)
                      (source2) edge [color = black,thick] (relay2)
                      (source3) edge [color = black] (relay1)
                      (source3) edge [color = black] (relay2)
                      (source1) edge [color = black] (dest1)
                      (source2) edge [color = black] (dest2)
                      (source3) edge [color = black] (dest3)
                      (relay1) edge [color = black,dashed] (dest1)
                      (relay1) edge [color = black,dashed] (dest2)
                      (relay1) edge [color = black,dashed] (dest3)
                      (relay2) edge [color = black,dashed] (dest1)
                      (relay2) edge [color = black,dashed] (dest2)
                      (relay2) edge [color = black,dashed] (dest3);

\end{tikzpicture}
\caption{Cooperative OFDM-based multi-source multi-destination mesh network with $K = 3$ and
$J = 2$.}
\label{fig:sysmodel}
\end{figure}
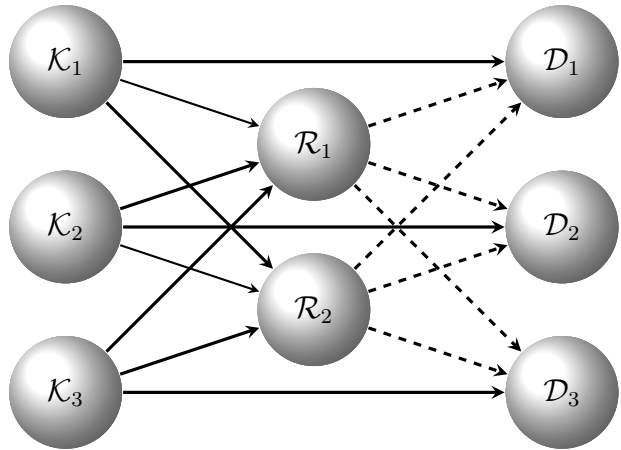

This paper considers an OFDM-based static mesh network of access points
(APs) as shown in Fig.~\ref{fig:sysmodel}. The network comprises $K$
sources assisted by $J$ dedicated relays. Each source node has its own
destination which is not within the set of sources and relays. Let
$\mathcal{K} = \left\{1,2,...,K\right\}$, $\mathcal{J} =
\left\{1,2,...,J\right\}$, and $\mathcal{N}_k =
\left\{1,2,...,N\right\}$ be the set of source nodes, relay nodes, and
subcarriers of source $k$, respectively. Note that the available bandwidth is
divided into $K\times N$ orthogonal subcarriers and each source node is assigned 
a block of OFDM and communicates over its corresponding band, i.e., \textit{simultaneous transmissions do not
interfere}. Here, for the sake of simplicity, we assume that all sources are allocated the same number of subcarriers. However,
the presented results can be easily generalized to the case of an unequal number of subcarriers per source node or to the case of OFDMA where multiple sources share an OFDM block. Furthermore, upon the admission of any new source node in the network, the subcarrier allocation needs to be updated. This, however, is beyond 
the scope of this paper. 

All sources and relays are attached to the power supply and transmit with constant and maximum
total power of $P$. We consider DF relaying wherein each relay
receives, decodes, and re-encodes the information with the same
codebook as the transmitter, and forwards it to the destination. Nodes
meet a half-duplex constraint. The
inter-node wireless channels are modeled as frequency-selective. Since
individual subcarriers of each source node experiences different
channel realizations, adaptive transmission strategy and implementing
power allocation at the sources and relays can enhance the system
performance. Finally, we assume that all inter-node channels vary
slowly enough for the channel state information (CSI) to be fed back to
a centralized unit with limited overhead, making resource allocation
possible.

The half duplex constraint imposes the need for a two-stage version of the DF
protocol. In Fig.~\ref{fig:sysmodel}, the solid arrows indicate the
first, time-sharing, stage wherein each source broadcasts its data
using $N$ subcarriers and \textit{each relay receives the OFDM block
from all source nodes on orthogonal channels}. During the second stage,
represented by dashed arrows, only those relays that can fully decode
the received information are nominated to participate. Finally, the
destination node combines messages received in the two phases to decode
the original information. This paper considers two cooperative
scenarios; treating each subcarrier as an independent transmission and,
more practically, cooperation at the level of an entire OFDM block.

The focus of this paper is to achieve max-min fairness in a
multi-source mesh network, i.e., to \textit{maximize the minimum rate across
all source nodes}. In the next section, we present resource allocation
schemes in such networks under the assumption that all relay nodes can
successfully decode received symbols.

\section{Resource Allocation with Ideal S-R Channels}
\label{sec:perfect}

This section develops optimal relay selection and power allocation in a
subcarrier-based fashion to achieve max-min fairness. In particular, two
block-based relaying schemes with different complexities are proposed.
The assumption of an ideal S-R channel, wherein a relay can always
decode the sources' transmission, is valid when the relays are close to
the sources or the S-R channels have a line-of-sight component. Note that this is a stepping stone to the next section where we drop this assumption.
Further, we assume that a source divides its power equally across all
subcarriers. 

\subsection{Subcarrier-Based Resource Allocation with Ideal S-R Channels}
In this section, each subcarrier is treated as an independent
transmission. In practice, the total number of subcarriers in a
network, $KN$, is much higher than the number of relay nodes, $J$.
Hence, a relay is most probably required to support multiple
subcarriers. In order to meet its power constraint, a relay must
distribute its available power amongst all subcarriers that it
supports. Under these conditions, the achievable rate of source $s_k$
over its $n^\mathrm{th}$ subcarrier is
\begin{eqnarray}
R_{k}^{(n)} & = & \max_{j} \min \left\{I_{s_kr_j}^{(n)},I_{s_kr_jd_k}^{(n)} \right\},
                                                                    \label{eq:MinOfTwo}\\
I_{s_kr_j}^{(n)} & = & \frac{1}{2}\log_{2}\left(1+\mathtt{SNR}_{kj}\alpha_{0k}^{(n)}
                        \lvert{h_{kj}^{(n)}}\rvert^2\right), \label{eq:SRrate} \\
\hspace{-2cm}I_{s_k r_j d_k}^{(n)} & = & \frac{1}{2}\log_{2}\left(1+
                                \mathtt{SNR}_{0k}\alpha_{0k}^{(n)}\lvert{h_{0k}^{(n)}}\rvert^2 +
            \mathtt{SNR}_{jk}\alpha_{jk}^{(n)}\lvert{h_{jk}^{(n)}}\rvert^2\right), \label{eq:SRDrate}
\end{eqnarray}
where $h_{0k}^{(n)}$, $h_{kj}^{(n)}$, and $h_{jk}^{(n)}$ denote the complex
S-D, S-R, and R-D channels over the $n^\mathrm{th}$ subcarrier of
$s_k$. $\mathtt{SNR}_{0k}$, $\mathtt{SNR}_{kj}$, and $\mathtt{SNR}_{jk}$
are the ratios of the total transmitted power to the power of noise;
 $\alpha_{0k}^{(n)}$ and $\alpha_{jk}^{(n)}$ are, respectively,
the fraction of the allocated power to the $n^\mathrm{th}$ subcarrier
of source $k$ at the source node and relay $j$.
Eqn.~\eqref{eq:MinOfTwo} declares that the rate of each source node
over its individual subcarriers is the minimum of the S-R rate
(Eqn.~\eqref{eq:SRrate}) and the compound S-R-D rate
(Eqn.~\eqref{eq:SRDrate}), i.e., the cooperative rate requires that
both the relay and destination fully decode the received data. The
total rate then is $R_k  = \sum_{\mathcal{N}_k} R_{k}^{(n)}$. Moreover, we assume that
\begin{itemize}
\item all S-R channels are ideal in that for any potential source
    rate
\begin{equation}
I_{s_{k}r_j}^{(n)} \geq I_{s_{k}r_jd_k}^{(n)} \hspace{0.2 cm} \Rightarrow
           \hspace{0.2 cm} R_{k}^{(n)} = I_{s_{k}r_jd_k}^{(n)}, \hspace*{0.2cm} \forall k,n.
           \nonumber
\end{equation}
\item source nodes distribute their available power equally amongst
    their subcarriers, i.e., $\alpha_{0k}^{(n)} = 1/N$. The power
    allocation problem is, therefore, relevant at the relays only.

\end{itemize}
In keeping with its many benefits described earlier, we impose a
selection constraint in the second, relaying, phase, i.e., each
subcarrier of a source node is relayed through \emph{at most} one of
the relays in the network. Therefore, the optimization problem we wish
to solve is
\begin{align}
&\max_{\alpha}\min_{k}{R_k} \label{eq:MaxMinI}\\
s.t. \hspace{0.2 cm}
&C_1:\alpha_{j_{1}k}^{(n)} \times \alpha_{j_{2}k}^{(n)}=0, \hspace{0.2 cm}
                \mbox{$\forall k,n,$ and $j_{1}\neq j_{2}$},\label{eq:SelectionI} \\
&C_2:\alpha_{jk}^{(n)}\geq {0},\hspace {0.2 cm} \forall {j,k,n}, \label{eq:NonNegI} \\
&C_3:\sum_{\mathcal{K}}\sum_{\mathcal{N}_k}\alpha_{jk}^{(n)} = 1,\hspace{0.2 cm} \forall {j},
                                                                    \label{eq:TotPowerI}
\end{align}
where $R_k = \sum_{\mathcal{N}_k}\frac{1}{2}\log_{2}\left(1+
\frac{1}{N}\mathtt{SNR}_{0k}\lvert{h_{0k}^{(n)}}\rvert^2
+\sum_{\mathcal{J}}
\mathtt{SNR}_{jk}\alpha_{jk}^{(n)}\lvert{h_{jk}^{(n)}}\rvert^2\right).$

Constraint $C_1$ enforces selection by allowing only one node to devote
power to each subcarrier. Constraints $C_2$ and $C_3$ state that the
amount of allocated power must be non-negative and that the total available power
of all relays is limited. Due to the selection
constraint,~\eqref{eq:MaxMinI}-\eqref{eq:TotPowerI} is an, essentially
intractable, mixed-integer programming optimization problem. One
proposed solution~\cite{SAE05, RC00} separates the power allocation and
selection problems. First, subcarriers are selected assuming equal
power allocation; then, power is distributed based on this selection.
However, with $K$ sources and $J$ relays, there are
$J^{KN}$ relay assignments to be checked. Therefore, even this scheme
is infeasible for realistic values of $K$, $J$, and $N$. We build on an
alternative approach developed in~\cite{KA09} to form an approximate
solution that is also an upper bound\footnote{It is worth emphasizing
that while the solution methodology here is similar to that
of~\cite{KA09}, both our problem formulation and solution are
significantly different. The development here, using the epigraph form,
leads to effective solutions to OFDM-based relaying and allows us to
show that selection is a \emph{sub-optimal} solution to the resource
allocation problem.}.

\subsubsection{An Approximate Solution and Upper Bound}
The objective function of the optimization problem is increasing and
concave in $\alpha_{jk}^{(n)}$. Other than the integer constraint
of~\eqref{eq:SelectionI}, the constraints in the original problem
of~\eqref{eq:MaxMinI}-\eqref{eq:TotPowerI} are convex. In order to find
an approximate, tractable, solution we first ignore the selection
constraint; hence, the solution to this modified
optimization problem will be an upper bound to the original
subcarrier-based (UBSB) resource allocation problem. The revised
formulation, stated here in \emph{epigraph form}, is a concave
maximization problem solvable in polynomial time using available
efficient solvers~\cite{BV04}.
\begin{align}
&\max_{\left\{t,\alpha\right\}} t \nonumber\\
s.t. \hspace{0.2 cm}
&C_1 : \sum_{\mathcal{N}_k}\frac{1}{2}\log_{2}\left(1+\frac{1}{N}\mathtt{SNR}_{0k}
        \lvert{h_{0k}^{(n)}}\rvert^2 +
     \sum_{\mathcal{J}}\mathtt{SNR}_{jk}\alpha_{jk}^{(n)}\lvert{h_{jk}^{(n)}}\rvert^2\right)-t\geq{0},
                                                    \hspace{0.2 cm} \forall{k},\nonumber\\
&C_2:\alpha_{jk}^{(n)}\geq0,\hspace{0.2 cm} \forall j,k,n, \hspace{0.5cm}
C_3:\sum_{\mathcal{K}}\sum_{\mathcal{N}_k}\alpha_{jk}^{(n)} = 1, \hspace
{0.2 cm} \forall {j}. \nonumber
\end{align}
The solution to this convex optimization problem is characterized by
the KKT conditions~\cite{BV04}. The Lagrangian is given by
\setlength{\arraycolsep}{0.0em}
\begin{eqnarray}
\mathcal{L}\left(\alpha_{jk}^{(n)},\gamma_{k},\mu_{j},\lambda_{jkn}\right) =t&+&\sum_{\mathcal{K}}
\gamma_{k}\left(\sum_{\mathcal{N}_k}\frac{1}{2}\log_{2}\left(1+\frac{1}{N}\mathtt{SNR}_{0k}
    \lvert{h_{0k}^{(n)}}\rvert^2 +\sum_{\mathcal{J}}\mathtt{SNR}_{jk}\alpha_{jk}^{(n)}
    \lvert{h_{jk}^{(n)}}\rvert^2\right)-t\right) \nonumber\\
&+&\sum_{\mathcal{J}} \mu_{j}\left(1-\sum_{\mathcal{K}}\sum_{\mathcal{N}_k}
\alpha_{jk}^{(n)}\right)+\sum_{\mathcal{J}}\sum_{\mathcal{K}}
            \sum_{\mathcal{N}_k}\lambda_{jkn}\alpha_{jk}^{(n)}, \nonumber
\end{eqnarray}
wherein $\gamma_{k}$, $\mu_{j}$, and $\lambda_{jkn}$ are Lagrangian multipliers.

For the sake of clarity, let us assume that $J = 2$, i.e., a
cooperative network comprising two relays. Since the problem is a
standard convex optimization problem and satisfies Slater's conditions,
any solution for the amount of power that relays $r_1$ and $r_2$
allocate to the $n^\mathrm{th}$ subcarrier of source $k$ satisfies the
KKT conditions. Suppose that both relays allocate some power to the $n^\mathrm{th}$ subcarrier of $s_k$.
Therefore, the convex problem satisfies the complementary slackness
condition, i.e., $\lambda_{r_{1}kn}=\lambda_{r_{2}kn}=0$. Now, we can
conclude that
$\frac{\mu_{1}}{\mu_{2}}=\frac{\rvert{h_{r_{1}k}^{(n)}}\rvert^{2}}
{\lvert{h_{r_{2}k}^{(n)}\rvert^{2}}}.$ Similarly, if the same two nodes
contribute to relaying the $i^\mathrm{th}$ subcarrier of source $k$,
using the same KKT conditions,
$\frac{\mu_{1}}{\mu_{2}}=\frac{\lvert{h_{r_{1}k}^{(i)}}\rvert^{2}}
{\lvert{h_{r_{2}k}^{(i)}}\rvert^{2}}. \label{eq:KKT2}$ These two
equations cannot be simultaneously satisfied since channel gains are
continuous random variables. Thus, \emph{at most one subcarrier} of each source
\emph{can} be helped by more than one relay.

Now, let us evaluate all the possible relay selections in a network
with $J = 3$, assuming the $n^\mathrm{th}$ subcarrier of source $k$ is
being helped by all relay nodes. The KKT conditions state that
$\frac{\mu_{1}}{\lvert{h_{r_{1}k}^{(n)}}\rvert^{2}} =
\frac{\mu_{2}}{\lvert{h_{r_{2}k}^{(n)}}\rvert^{2}} =
\frac{\mu_{3}}{\lvert{h_{r_{3}k}^{(n)}}\rvert^{2}}.
\label{eq:KKT3Nodes1}$ Now suppose that the $i^\mathrm{th}$ subcarrier
of the same source is relayed through $r_{1}$ and $r_{2}$, i.e.,
$\frac{\mu_{1}}{\lvert{h_{r_{1}k}^{(i)}}\rvert^{2}} =
\frac{\mu_{2}}{\lvert{h_{r_{2}k}^{(i)}}\rvert^{2}}.
\label{eq:KKT3Nodes2}$ Thus,  we have
$\lvert{h_{r_{1}k}^{(n)}}\rvert^{2}/\lvert{h_{r_{2}k}^{(n)}}\rvert^{2}
=
\lvert{h_{r_{1}k}^{(i)}}\rvert^{2}/\lvert{h_{r_{2}k}^{(i)}}\rvert^{2},$
which is a zero-probability event.

Consider, again, the scenario in which none of the subcarriers can be
helped with all three relays. As an example, consider the case where
the $n^\mathrm{th}$ subcarrier is relayed via nodes $r_1$ and $r_2$ and
the $i^\mathrm{th}$ subcarrier is helped by node $r_1$ and $r_3$.
Applying the same KKT conditions, it follows that
$\frac{\mu_{1}}{\lvert{h_{r_{1}k}^{(n)}}\rvert^{2}}
=\frac{\mu_{2}}{\lvert{h_{r_{2}k}^{(n)}}\rvert^{2}} \hspace*{0.1in}
\mathrm{and} \hspace*{0.1in}
\frac{\mu_{1}}{\rvert{h_{r_{1}k}^{(i)}}\rvert^{2}} =
\frac{\mu_{3}}{\lvert{h_{r_{3}k}^{(i)}}\rvert^{2}}.$ Now, the
$m^\mathrm{th}$ subcarrier can be helped by node $r_2$ and $r_3$ only
if $\frac{\lvert{h_{r_{2}k}^{(n)}}\rvert^{2}}
{\lvert{h_{r_{1}k}^{(n)}}\rvert^{2}\lvert{h_{r_{2}k}^{(m)}}\rvert^{2}}=
\frac{\lvert{h_{r_{3}k}^{(i)}}\rvert^{2}}{\lvert{h_{r_{1}k}^{(i)}}\rvert^{2}
\mid{h_{r_{3}k}^{(m)}}\rvert^{2}}$, which happens with zero
probability. Therefore, when two subcarriers are relayed by two nodes,
\emph{all others can be helped by at most one node.}

Generalizing this to the network with $K$ sources and $J$ relays,
\emph{one concludes that at most $J-1$ subcarriers of each source can
be helped by more than one relay and selection is imposed on $(N-J+1)$
subcarriers.} In practice, $N\gg J$ which means that a large fraction
of subcarriers meet the selection criterion, i.e., selection is an
\emph{approximate, though not optimal solution}, to the relaxed
optimization problem. Note that this contradicts the work
in~\cite{PNB08} which suggests that selection is \emph{optimal}.

\subsubsection{ A Heuristic Algorithm and a Lower Bound}
\label{sec:HeuristicPerfectSR}

By neglecting the selection constraint, the solution to the modified
problem provides an upper bound to that of the original optimization
problem in~\eqref{eq:MaxMinI}-\eqref{eq:TotPowerI}. Here, we use this
to develop a heuristic solution to the original problem. We force the
subcarriers that do not meet the constraint (a maximum of $J-1$ of
them) to receive power only from the single relay that achieves a
higher data rate. Mathematically speaking
\begin{align}
&r_{k}^{(n)}  =  r_{m}^{(n)}, \hspace{0.5cm}
m  =  \arg \max_{j} \left\{\frac{1}{2}\log_{2}\left(1+\frac{1}{N}\mathtt{SNR}_{0k}
                \lvert{h_{0k}^{(n)}}\rvert^2 +\mathtt{SNR}_{jk}\alpha_{jk}^{(n)}
                                \lvert{h_{jk}^{(n)}}\rvert^2\right)\right\}, \nonumber
\end{align}
where $r_{k}^{(n)}$ is the relay node which contributes to the
transmission of source $k$ on the $n^\mathrm{th}$ subcarrier. Since
this solution meets all the constraints of the original problem, this
is also a \emph{lower bound} to the solution of the subcarrier based
(LBSB) optimization problem. In Section~\ref{sec:Simulation}, we will
show that the upper and lower bounds are indistinguishable. As a
result, this heuristic approach provides almost the exact solution to
the original mixed-integer optimization problem with significantly
reduced solution complexity.

\subsection{Block-Based Resource Allocation with Ideal S-R Channels}\label{sec:Perf_BB}
The optimization problem and solution detailed so far is in keeping
with existing literature. It allows different subcarriers within an
OFDM block to be helped by different relays. This is problematic for
two reasons. One, while not explicitly stated, most of the previous
work assumes a relay can treat each subcarrier as an independent
transmission. In DF-based relaying, the decoding constraint is at the
level of a subcarrier, e.g., \eqref{eq:MinOfTwo}. However, in OFDM, the
data is first protected by a channel code, modulated and then a block
of $N$ subcarriers is formed. It is not possible to decode information
without receiving and decoding an entire OFDM block. Second, practical
OFDM systems depend heavily on accurate time and frequency
synchronization. This would be extremely difficult in a distributed
mesh network.

In a multi-source network, as long as each relay has to divide its
available power amongst all allocated sources, the solution to the
relay assignment problem is not immediate. Here, we separate the
problem into selection followed by power allocation (via waterfilling)
across subcarriers. As in~\cite{BA08}, two selection schemes
with different levels of complexity are proposed and results will be
compared in terms of the max-min rate in Section~\ref{sec:Simulation}.

\subsubsection{Optimal Relay Selection}
In a network with $K$ sources and $J$ relays, there are $J^K$
different possible relay assignments. The optimal scheme is exhaustive
search over all possible relay selections and pick the one which
provides the maximal minimum rate. This is clearly impossible for any
reasonable $K$ and $J$.

\subsubsection{Decentralized Relay Selection}\label{sec:Dec_Relay_Sel}
The decentralized or simple relay selection scheme ignores all other
sources. Each source chooses its best relay with the assumption that
the corresponding relay distributes its power equally over all
subcarriers of \emph{only that source}. In particular
\begin{align}
r_k&=r_m, \hspace{0.5cm}
m=\arg \max_{j} \left(\sum_{\mathcal{N}_k}\log_2\left( 1+\frac{1}{N}\mathtt{SNR}_{jk}
                                \lvert{h_{jk}^{(n)}}\rvert\right)\right). \nonumber
\end{align}
With each source having selected the relays, power is allocated via waterfilling, to the
assigned sources. Note that since each source-destination pair only needs local CSI and
selection is performed independently of all other sources, this scheme
can be implemented in a decentralized manner. In a network with $J$
dedicated relays, only $J$ water-filling problems need to be solved.

\section{Resource Allocation with Finite-Power S-R Channels}
\label{sec:imperfect}

The previous section developed solutions under the assumption of an
ideal S-R channel. In this section we consider the general case of
resource allocation across the S-D, S-R, and R-D channels. The solution
to this optimization problem also chooses the best transmission
strategy for each source, i.e., direct transmission is a valid solution
if that is optimal. Our approach also allows us to move beyond
heuristics for block-based relaying.

\subsection{Subcarrier-Based Resource Allocation with Finite-Power S-R Channels}
\label{sec:subimper}

Given the fact that each source is allowed to switch between DF and
direct transmission, one concludes that in a network with $K$ sources
and $J$ relays, with subcarrier based relaying
\begin{align}
R_{k}^{(n)}  =  \max\left\{I_{s_kd_k}^{(n)},\max_{j} \min \left\{I_{s_kr_j}^{(n)},
                            I_{s_kr_jd_k}^{(n)} \right\}\right\}, \label{eq:MinOfThree}
\end{align}
wherein $I_{s_kd_k}^{(n)}  =
\log_{2}\left(1+\mathtt{SNR}_{0k}\alpha_{0k}^{(n)}\lvert{h_{0k}^{(n)}}\rvert^2\right)$.
Eqn.~\eqref{eq:MinOfThree} declares that the rate of each source node
over its individual subcarriers is the maximum of the direct and
cooperative transmission rates; in turn, the cooperative rate requires
that both the relay and destination fully decode the received data. The
total rate is then the sum of achievable rates of all subcarriers. In
addition, by taking the S-R channel into account, optimal power
allocation at the source nodes may alter relay selection and further
enhance the performance of the network.

Let $\mathcal{J}_+ = \left\{0,1,2,...,J\right\}$ be an extended set of
relays; the `0' index indicates direct transmission. Therefore, the
formal optimization problem is
\begin{align}
&\max_{\alpha}\min_{k}{R_k} \label{eq:MaxMinII}\\
s.t. \hspace{0.2 cm}
&C_1:\alpha_{j_{1}k}^{(n)} \times \alpha_{j_{2}k}^{(n)}=0, \hspace{0.1 cm}
                            \forall k,n, \left\{j_1,j_2\right\} \in \mathcal{J},
\label{eq:SelectionII} \\
&C_2:\alpha_{jk}^{(n)}\geq {0},\hspace {0.1 cm} \forall k,n,j\in \mathcal{J}_+,
                                                                \label{eq:NonNegII} \\
&C_3:\sum_{\mathcal{K}}\sum_{\mathcal{N}_k}\alpha_{jk}^{(n)} \leq 1,\hspace{0.1 cm}
                            \forall j \in \mathcal{J}, \label{eq:TotPowerII}\\
&C_4:\sum_{\mathcal{N}_k}\alpha_{0k}^{(n)} = 1, \hspace{0.1 cm} \forall {k},
\label{eq:TotPowerII_Source}
\end{align}
Eqns.~\eqref{eq:SelectionII}-\eqref{eq:NonNegII} are equivalent to the
constraints \eqref{eq:SelectionI}-\eqref{eq:TotPowerI} of the original
optimization problem of the previous section. Unlike the previous
optimization problem, since source nodes are allowed to transmit
directly, some relays might stay silent during the second time-slot.
Hence, as stated in Eqn.~\eqref{eq:TotPowerII}, the power constraint
can be satisfied by inequality. Eqn.~\eqref{eq:TotPowerII_Source}
limits the available power of each source node. Similar to the previous
scenario, since each relay must split its available power amongst all
source nodes which it supports, transmission strategy selection, relay
assignment, and power allocation problem is combinatorial and needs to
be solved jointly.

\subsubsection{An Approximate Solution and Upper Bound}
To make the problem mathematically tractable, we introduce
$KN(J+1)$
indicator variables to the objective function. Therefore, the new
optimization problem can be expressed as
\begin{align}
&\max_{\left\{\alpha,\rho \in \left\{0,1\right\}\right\}}
                                                            \min_{k}{R_k} \nonumber\\
s.t. \hspace{0.2 cm}
&C_1:\alpha_{jk}^{(n)}\geq {0},\hspace {0.1 cm}\rho_{jk}^{(n)}\geq {0}, \hspace{0.1 cm}
            \forall k,n,j\in \mathcal{J}_+,\nonumber \hspace{0.5cm}
            C_2:\sum_{\mathcal{K}}\sum_{\mathcal{N}_k}\rho_{jk}^{(n)}\alpha_{jk}^{(n)} \leq 1,
            \hspace{0.1 cm} \forall j \in \mathcal{J}, \nonumber\\
&C_3:\sum_{\mathcal{J}_+}\sum_{\mathcal{N}_k}\rho_{jk}^{(n)}\alpha_{0k}^{(n)} = 1,
                                                    \hspace{0.1 cm} \forall {k},
\hspace{2.25cm}C_4:\sum_{\mathcal{J}_+}\rho_{jk}^{(n)}=1, \hspace{0.1 cm} \forall{k,n}. \nonumber
\end{align}
If  source $k$ allocates a
fraction of its available power to the $n^\mathrm{th}$ subcarrier, i.e,
$\alpha_{0k}^{(n)} \neq 0$, for any set of $\alpha_{jk}^{(n)}$
satisfying \eqref{eq:SelectionII}-\eqref{eq:TotPowerII}, the following equation is true
\begin{equation}
\rho_{jk}^{(n)} = \left\{
\begin{array}{rl}
1, & \hspace{0.1cm}\alpha_{jk}^{(n)} \neq 0,\\
0, & \hspace{0.1cm}\alpha_{jk}^{(n)} = 0.
 \label{eq:rho_selection}
\end{array} \right.
\end{equation}
Eqn.~\eqref{eq:rho_selection}, along with the fact that only one
indicator variable of each source can be non-zero at a time, enforces
the selection constraint of the original problem. The total rate of $s_k$ is
\begin{align}
R_k &= \sum_{\mathcal{N}_k}\rho_{0k}^{(n)}\log_2\left(1+\mathtt{SNR}_{0k}\alpha_{0k}^{(n)}
        \lvert{h_{0k}^{(n)}}\rvert^2\right) + \nonumber \\
        & \sum_{\mathcal{J}}\sum_{\mathcal{N}_k}\rho_{jk}^{(n)}\min\left\{
        \frac{1}{2}\log_2\left(1 + \mathtt{SNR}_{kj}\alpha_{0k}^{(n)}
        \lvert{h_{kj}^{(n)}}\rvert^2\right),\frac{1}{2}\log_2\left(1 +
        \mathtt{SNR}_{0k}\alpha_{0k}^{(n)}\lvert{h_{0k}^{(n)}}\rvert^2 +
        \mathtt{SNR}_{jk}\alpha_{jk}^{(n)}\lvert{h_{jk}^{(n)}}\rvert^2\right)\right\}. \nonumber
\end{align}
Note that, since indicator variables can only take integer values, the
problem is still a combinatorial optimization problem. As in the
previous section, our solution methodology is to relax the
corresponding constraint and allow each stream to be transmitted both
directly as well as cooperatively through multiple relays. Thus,
indicator variables of each individual subcarriers can take any
rational value on the convex hull of the original discrete set.
Consequently, the resulting solution is an
upper bound to the min-rate of the original subcarrier-based problem (UBSB) formulated
in~\eqref{eq:MaxMinII}-\eqref{eq:TotPowerII_Source}. Furthermore,
$\rho_{jk}^{(n)}$ can now be interpreted as a \emph{fraction of time}
that $s_k$ transmits over its $n^\mathrm{th}$ subcarrier directly
($j=0$) and cooperatively ($j\in \mathcal{J}$).

The rate $R_k$ comprises three different terms: S-D, S-R, and S-R-D
rates. One can easily show that none of them is jointly concave in the
set of variables. By a change of variables as in~\cite{WCLM99}, we set
\begin{align}
&\rho_{jk}^{(n)}\alpha_{jk}^{(n)} = r_{jk}^{(n)}, \hspace{0.1cm} j \in \mathcal{J}_+,
\hspace{0.5cm}
\rho_{jk}^{(n)}\alpha_{jk}^{(n)} = p_{jk}^{(n)}, \hspace{0.1cm} j \in \mathcal{J}.
\nonumber
\end{align}
It is worth noting that this is a key difference from the work
in~\cite{LYLGL08,LLWLL09} which did not take the coupling constraint
between time-sharing coefficients and power allocation into account.

The new optimization problem in terms of $(\rho,r,p)$ can be formulated
as
\begin{align}
&\max_{\left\{\rho,r,p\right\}}\min_{k}{R_k}
\label{eq:MaxMinIV}\\
s.t. \hspace{0.2 cm}
&C_1:\rho_{jk}^{(n)}\geq r_{jk}^{(n)}\geq 0,\hspace {0.1 cm} \forall k,n,j\in \mathcal{J}_+,
\hspace{0.5cm} C_2:\rho_{jk}^{(n)}\geq p_{jk}^{(n)}\geq 0, \hspace{0.1 cm} \forall k,n,j\in
\mathcal{J},
\label{eq:C1} \\
&C_3:\sum_{\mathcal{K}}\sum_{\mathcal{N}_k}p_{jk}^{(n)} \leq 1,\hspace{0.1 cm}
\forall j \in \mathcal{J}, \hspace{1.55cm}C_4:\sum_{\mathcal{J}_+}
\sum_{\mathcal{N}_k}r_{jk}^{(n)} = 1, \hspace{0.1 cm} \forall {k},
\label{eq:C4}\\
&C_5:\sum_{\mathcal{J}_+}\rho_{jk}^{(n)}=1, \hspace{0.1 cm} \forall{k,n}\label{eq:C5}.
\end{align}
Hence, the achievable rate of source $k$ is expressed as
\begin{align}
R_k &= \sum_{\mathcal{N}_k}\rho_{0k}^{(n)}\log_2\left(1+\frac{\mathtt{SNR}_{0k}r_{0k}^{(n)}
        \lvert{h_{0k}^{(n)}}\rvert^2}{\rho_{0k}^{(n)}}\right) + \nonumber \\ &
        \sum_{\mathcal{J}}\sum_{\mathcal{N}_k}\rho_{jk}^{(n)}
        \min\left\{\frac{1}{2}\log_2\left(1 + \frac{\mathtt{SNR}_{kj}r_{jk}^{(n)}
        \lvert{h_{kj}^{(n)}}\rvert^2}{\rho_{jk}^{(n)}}\right),
        \frac{1}{2}\log_2\left(1 + \frac{\mathtt{SNR}_{0k}r_{jk}^{(n)}
        \lvert{h_{0k}^{(n)}}\rvert^2}{\rho_{jk}^{(n)}} + \frac{\mathtt{SNR}_{jk}p_{jk}^{(n)}
        \lvert{h_{jk}^{(n)}}\rvert^2}{\rho_{jk}^{(n)}}\right)\right\}. \nonumber
\end{align}
\vspace*{2ex}
\newtheorem{theorem}{\indent Theorem}
\begin{theorem}
The objective function in~\eqref{eq:MaxMinIV} is jointly concave in
$\rho$, $r$, and $p$.
\end{theorem}
\begin{IEEEproof}
The S-D and S-R rates are in the form of $f(x,y) = x\log\left(1
+y/x\right)$ and the rate of the compound S-R-D channel is in the form
of $g(x,y,z) =x\log\left(1+y/x+z/x\right)$. In addition, $x$, $y$, and
$z$ are non-negative variables. One can show that the Hessian of $f$ is
\begin{equation}
\nabla^{2}f=\frac{1}{(1+y/x)^2}
\left[
\begin{array}{cc}
-y^2/x^3 & y/x^2 \\
y/x^2 & -1/x
\end{array} \right] \nonumber.
\end{equation}
The determinant of $\nabla^{2}f$, the product of the eigenvalues, is
zero. The trace of the $\nabla^{2}f$, the sum of the eigenvalues, is a
negative value, which certifies that the $\nabla^{2}f\preceq 0$, i.e.,
the Hessian evaluated within the optimization region is a negative
semi-definite matrix. Now, let us follow the same strategy to show that
the third term is also jointly concave in the set of consisting
variables. Therefore
\begin{equation}
\nabla^{2}g=\frac{1}{(1+y/x+z/x)^2}
\left[
\begin{array}{lcr}
-(y+z)^2/x^3 \hspace{0.2cm}&  (y+z)/x^2 \hspace{0.2cm} & (y+z)/x^2 \\
(y+z)/x^2 & -1/x & -1/x \\
(y+z)/x^2 & -1/x & -1/x
\end{array} \right]. \nonumber
\end{equation}
Similar to the previous case, the determinant and trace of the
$\nabla^{2}g$ are, respectively, zero and negative. Moreover,
$\nabla^{2}g$ is a rank one matrix, i.e., it has one non-positive
eigenvalue and two zero ones. Thus, $\nabla^{2}g$ is a negative
semi-definite matrix which proves that the rate of the S-R-D channel is
jointly concave in $\left(\rho,r,p\right)$. It is also known that a
point-wise minimum and the non-negative summation of a set of concave
functions are also concave functions~\cite{BV04}. Hence, the underlying
objective function is jointly concave in $\left(\rho,r,p\right)$.
\end{IEEEproof}
Although the objective function is jointly concave, it is not
differentiable. By rewriting it in the \emph{epigraph form}, the final
optimization problem can be stated as follows
\begin{align}
\hspace{-0.5cm}&\max_{\left\{t,\zeta,\rho,r,p\right\}} t \nonumber\\
s.t. \hspace{0.2cm}\nonumber
&C_1: \eqref{eq:C1}-\eqref{eq:C5}\hspace{0.3cm} \nonumber \\
& C_2:\sum_{\mathcal{J}_+}
            \sum_{\mathcal{N}_k}\zeta_{jk}^{(n)} \geq t, \hspace{0.05cm} \forall k,\\
&C_3:\rho_{0k}^{(n)}\mathcal{C}\left(\frac{\mathtt{SNR}_{0k}r_{0k}^{(n)}
        \lvert{h_{0k}^{(n)}}\rvert^2}{\rho_{0k}^{(n)}}\right) \geq \zeta_{0k}^{(n)},
                                                    \hspace{0.1cm} \forall k,n, \nonumber\\
&C_4:\frac{\rho_{jk}^{(n)}}{2}\mathcal{C}\left(\frac{\mathtt{SNR}_{kj}r_{jk}^{(n)}
        \lvert{h_{kj}^{(n)}}\rvert^2}{\rho_{jk}^{(n)}}\right)\geq \zeta_{jk}^{(n)},
        \hspace{0.1cm} \forall k,n,j \in \mathcal{J},\nonumber\\
&C_5:\frac{\rho_{jk}^{(n)}}{2}\mathcal{C}\left(\frac{\mathtt{SNR}_{0k}r_{jk}^{(n)}
            \lvert{h_{0k}^{(n)}}\rvert^2+ \mathtt{SNR}_{jk}p_{jk}^{(n)}
            \lvert{h_{jk}^{(n)}}\rvert^2}{\rho_{jk}^{(n)}}\right)
                \geq \zeta_{jk}^{(n)}, \hspace{0.1cm} \forall k,n,j \in \mathcal{J},\nonumber
\end{align}
where $\mathcal{C}(x) = \log_2(1 + x)$. This modified version is a
standard convex optimization problem which, again, can be solved using
well established and efficient iterative algorithms~\cite{BV04}.

\subsubsection{A Heuristic Algorithm and a Lower Bound}\label{sec:lbsb}
The upper bound derived in the previous section approximates, but does
not meet the selection constraint. As in
Section~\ref{sec:HeuristicPerfectSR}, our approach to imposing
selection is to assign to each subcarrier the transmission strategy and
the relay that provides the maximum achievable rate. Thus, the selection
constraint is enforced as
\begin{align}
R_k^{(n)} & = \max \left\{I_{s_kd_k}^{(n)},\min\left\{I_{s_kr_m}^{(n)},
                                I_{s_kr_md_k}^{(n)}\right\}\right\},\hspace{0.5cm}
m  =\arg \max_{j} \min \left\{I_{s_kr_j}^{(n)},I_{s_kr_jd_k}^{(n)}\right\}. \nonumber
\end{align}
Since this solution satisfies all constraints of the original
problem in~\eqref{eq:MaxMinII}-\eqref{eq:TotPowerII_Source}, this
heuristic scheme provides a \emph{lower bound} (LBSB). Moreover, the power
freed up by the selection step can be reused by waterfilling over other
source nodes which are helped by each individual relays. However, as we
show in Section~\ref{sec:Simulation}, the performance gap is not
noticeable; thus, there is no need to apply a second round of
waterfilling. Finally, it is worth emphasizing that if direct
transmission were optimal, the power allocated at all relays would be
zero, i.e., \emph{the approach is adaptive across relay strategies.}

\subsection{Block-Based Resource Allocation with Finite-Power S-R Channels}
This section deals with the selection and power allocation at the level
of an entire OFDM block in the general case of finite-power S-D, S-R and
R-D channels. As in the previous section, the solution to this problem
optimizes the transmission strategy for each individual source node.
The achievable rate of each source node across the entire OFDM block is
\begin{equation}
R_{k} = \max\left\{\sum_{\mathcal{N}_k}I_{s_kd_k}^{(n)},\max_{j} \min
\left\{\sum_{\mathcal{N}_k}I_{s_kr_j}^{(n)},\sum_{\mathcal{N}_k}I_{s_kr_jd_k}^{(n)}
\right\}\right\}, \nonumber
\end{equation}
which states that each OFDM block can be transmitted either directly
or via the relay node which supports a higher data rate. Here, unlike
in~\eqref{eq:MinOfThree}, the individual terms in the rate expression
include a sum over all subcarriers; in block-based selection, all
subcarriers are relayed by the same relay node. The formal optimization
problem is therefore
\begin{align}
&\max_{\alpha}\min_{k}{R_k} \nonumber\\
s.t. \hspace{0.2 cm}
&C_1:\sum_{\mathcal{N}_k}\alpha_{j_{1}k}^{(n)} \times
        \sum_{\mathcal{N}_k}\alpha_{j_{2}k}^{(n)}=0, \hspace{0.1 cm}
        \forall k, \left\{j_1,j_2\right\} \in \mathcal{J}, \nonumber\\
&C_2:\alpha_{jk}^{(n)}\geq {0},\hspace {0.1 cm} \forall k,n,j\in
    \mathcal{J}_+ \nonumber, \nonumber \\
&C_3: \sum_{\mathcal{K}}\sum_{\mathcal{N}_k}\alpha_{jk}^{(n)} \leq 1,\hspace{0.1 cm}
                                                \forall j \in \mathcal{J}, \hspace*{1.5cm}
 C_4:\sum_{\mathcal{N}_k}\alpha_{0k}^{(n)} = 1, \hspace{0.1 cm} \forall {k}.
\nonumber
\end{align}
$C_1$ states that each OFDM \emph{block} can be helped by at most one
relay node. Other constraints are similar to those of the original
subcarrier-based scheme formulated in Section~\ref{sec:subimper}.

\subsubsection{An Approximate Solution and Upper Bound}
The block-level optimization problem is an integer programming with
exponential complexity. Thus, again, we introduce time-sharing
coefficients to the objective function and rewrite the achievable rate
of source $k$ as
\begin{align}
R_k &= \rho_{0k}\sum_{\mathcal{N}_k}\log_2\left(1+\mathtt{SNR}_{0k}\alpha_{0k}^{(n)}
\lvert{h_{0k}^{(n)}}\rvert^2\right) + \nonumber \\ & \hspace{-0.2cm}
\sum_{\mathcal{J}}\rho_{jk}\min\left\{\frac{1}{2}
\sum_{\mathcal{N}_k}\log_2\left(1 + \mathtt{SNR}_{kj}\alpha_{0k}^{(n)}
\lvert{h_{kj}^{(n)}}\rvert^2\right),\sum_{\mathcal{N}_k}\frac{1}{2}
\log_2\left(1 + \mathtt{SNR}_{0k}\alpha_{0k}^{(n)}\lvert{h_{0k}^{(n)}}\rvert^2 +
\mathtt{SNR}_{jk}\alpha_{jk}^{(n)}\lvert{h_{jk}^{(n)}}\rvert^2\right)\right\}. \nonumber
\end{align}
Note that, since relaying is block-based, only $K(J+1)$ time-sharing
coefficients are required. Following the same approach as the previous
section, we relax the selection constraint and set
\begin{align}
&\rho_{jk}\alpha_{0k}^{(n)} = r_{jk}^{(n)}, \hspace{0.1cm} j \in \mathcal{J}_+,\hspace{0.5cm}
\rho_{jk}\alpha_{jk}^{(n)} = p_{jk}^{(n)}, \hspace{0.1cm} j \in \mathcal{J}. \nonumber
\end{align}
Using Theorem 1, it is straight forward to prove that the resulting
optimization problem is jointly concave in $\left(\rho,r,p\right)$.
Finally, by rewriting the objective function in epigraph form, a
standard convex optimization problem can be formulated. Since we
relaxed the selection constraint, this solution provides an upper bound
to the minimum rate of the original block-based relaying (UBBB). The
approach developed in the Section~~\ref{sec:subimper} therefore
provides the basis for block-based optimization as well.

\subsubsection{A Heuristic Algorithm and a Lower Bound}
Having generalized our approach in Section~\ref{sec:lbsb} to the
block-based relaying, the lower bound to the block-based scheme (LBBB)
can be achieved by choosing the best relay for individual source nodes.
\subsubsection{Decentralized Resource Allocation}
The  solution to the optimization problem detailed so far requires
joint selection of the transmission strategy, the relay node, and power
allocation to each source in the network. This solution requires a
central resource allocation unit which has the full CSI of all
channels. This requires significant transmission and coordination
overhead, potentially making the solution impractical. In this section,
we develop a simplified decentralized scheme, wherein, similar
to~\ref{sec:Dec_Relay_Sel}, at the first stage each source selects its
best relay independently
\begin{align}
r_k&=r_m, \hspace{0.2cm}  \hspace{0.5cm} m=\arg \max_{j}
\left\{\min\left\{\sum_{\mathcal{N}_k}I_{s_kr_j}^{(n)},
            \sum_{\mathcal{N}_k}I_{s_kr_jd_k}^{(n)}\right\}\right\}. \nonumber
\end{align}
Second, the transmission strategy is chosen by comparing the rates of
the direct and relaying transmissions. Given that each individual
source node has already selected its transmission strategy, \emph{at
most} $J$ waterfilling problems need to be solved to maximize the
minimum rate across source nodes. Furthermore, if a source node has
decided to transmit directly, power is distributed based on the S-D
channel state. In Section~\ref{sec:Simulation}, we will show that, in
fact, the performance of the distributed scheme closely tracks that of
the UBBB algorithm.

\section{Simulation Results and Discussion} \label{sec:Simulation}

This section presents simulation results for the proposed relay
selection and resource allocation schemes described in
Sections~\ref{sec:perfect} and~\ref{sec:imperfect}. We consider two
different network geometries. In the first scenario, all inter-node
channels are modeled as independent and identically distributed
(i.i.d.) random variables. The second network setup is more realistic;
in that nodes are randomly distributed. The
communication channels are modeled using the COST-231 channel model
recommended by the IEEE 802.16j working group~\cite{COST}. The chosen
parameters for the COST-231 are given in Table~\ref{tb:tabel1}.

\subsection{Resource Allocation with Infinite-Power S-R Channels in I.I.D.
Scenario}

\begin{figure}[!t]
\centering
\includegraphics[width=11cm,height=8cm]{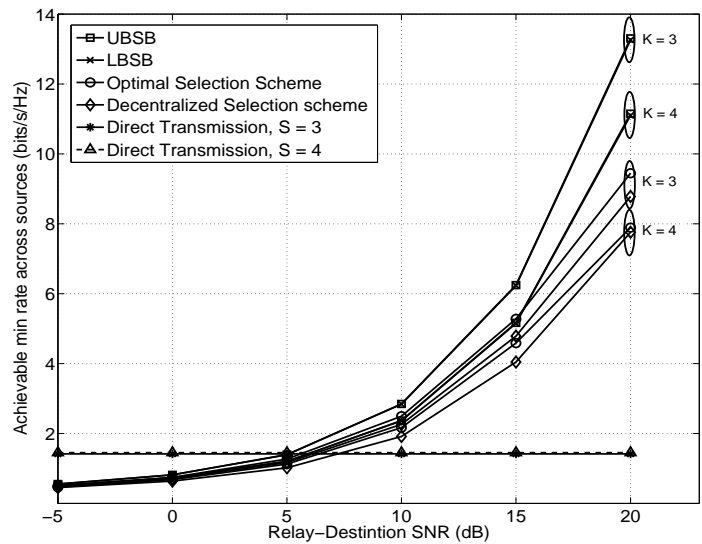}
\caption{Achievable min rate of different resource allocation strategies in ``i.i.d channel"
                    scenario with $J=2$ and $N=32$.}
\label{fig:MaxMin_IID_Perf}
\end{figure}

Our first example implements relay selection and resource allocation
for a mesh network with $K=3$ and $K=4$ APs with $N=32$. The S-R channels are assumed ideal; the average SNR of all
S-D channels is set to $5$dB. Fig.~\ref{fig:MaxMin_IID_Perf} plots the
minimum achievable rate across the $K$ source nodes for different
values of the R-D SNRs. As can be seen from the figure, the upper and
lower bounds are indistinguishable. In this setup, \emph{at most one}
of the subcarriers of each source node can be helped by both relays.
Since the number of subcarriers, $N$, is generally much larger than the
number of relays, $J$, selection is close-to-optimal.

Given the additional flexibility of subcarrier-based cooperation
schemes, both UBSB and LBSB outperform block-based schemes. Moreover,
although the optimal block-based relaying scheme is computationally
much more complex than the decentralized scheme, the performance
benefit is negligible. Enforcing direct transmission has the worst
performance, validating the fact that cooperation transmission can
boost network performance under the max-min metric.

\subsection{Resource Allocation with Infinite-Power S-R Channels in Distributed
Scenario}

In this example, nodes are geographically distributed and inter-node
channels are modeled using the COST-231 channel model. We generate the
random node locations over an square area of 200m $\times$ 200m. Source
and destination nodes are located on the edges of the square area while
relays are randomly distributed within the area. The variance of the
log-normal fading is set to $10.6$dB. In this experiment, for each set
of locations, independent channel realizations are simulated and
results averaged over both node locations and channel states.

\begin{figure}[!t]
\centering
\includegraphics[width=11cm,height=8cm]{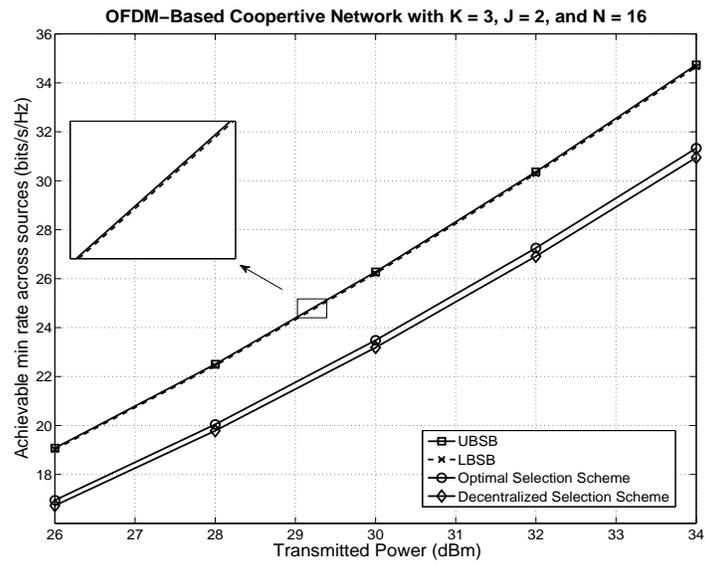}
\caption{Achievable min rate of different resource allocation strategies in ``distributive"
 scenario with $J=2$ and $N=16$.}
\label{fig:MaxMin_Dist_Perf}
\end{figure}

Fig.~\ref{fig:MaxMin_Dist_Perf} plots the max-min achievable rate
across all APs and compares the performance of various resource
allocation schemes. From the figure, the performance gap between UBSB
and LBSB is, again, negligible; the heuristic method to find the
solution of the original convex optimization problem is almost exact.
However, it is worth emphasizing that in both
Figs.~\ref{fig:MaxMin_IID_Perf} and~\ref{fig:MaxMin_Dist_Perf}
\emph{there is} a difference, albeit minuscule, between the UBSB and
LBSB performance. This proves the fact that selection, is an
\emph{approximate, not optimal} solution.

Fig.~\ref{fig:MaxMin_Dist_Perf} also compares the performance of
block-based schemes. Simple relay selection closely tracks the optimal
relay selection method, but with significantly less complexity. This
result indicates that the simple relay selection scheme can be
implemented in a decentralized manner without significant performance
loss.

\begin{figure}[!t]
\centering
\includegraphics[width=11cm,height=8cm]{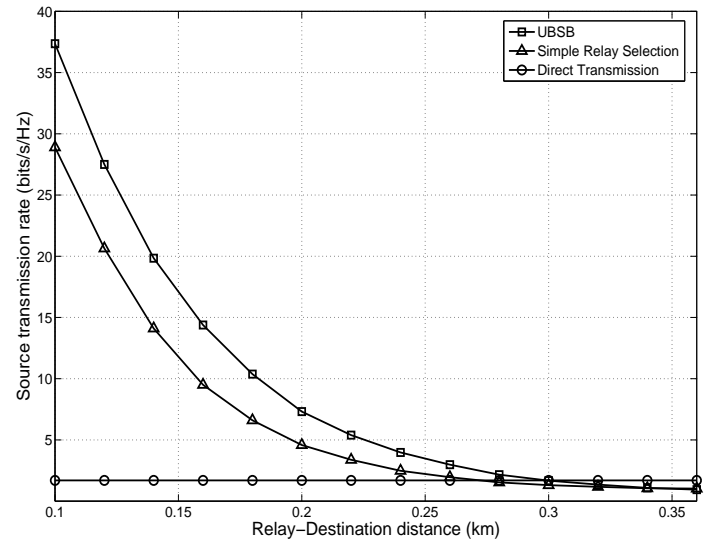}
\caption{Source transmission rate in a single source-destination pair network with $J=2$ and
$N=16$.}
\label{fig:TwoNodes}
\end{figure}

Fig.~\ref{fig:TwoNodes} illustrates the importance of node locations
on the performance of different resource allocation algorithms. This
example simulates a single source-destination pair with two relay
nodes. The S-D distance is fixed to $0.2\sqrt{2}$ km. Relays are
located on both sides of the S-D path. Clearly one wants to use the relay
close to the destination; however, note that this may impact the
assumption that relays can always decode. Simulation results show
that relaying schemes outperform direct transmission whenever relays
are located between the source and destination nodes. While the upper
bound on subcarrier-based selection outperforms block-based selection,
the performance loss for this more practical approach is surprisingly
small.

\subsection{Resource Allocation with Finite-Power S-R Channels in I.I.D.
Scenario} 
\begin{figure}[!t]
\centering
\includegraphics[width=11cm,height=8cm]{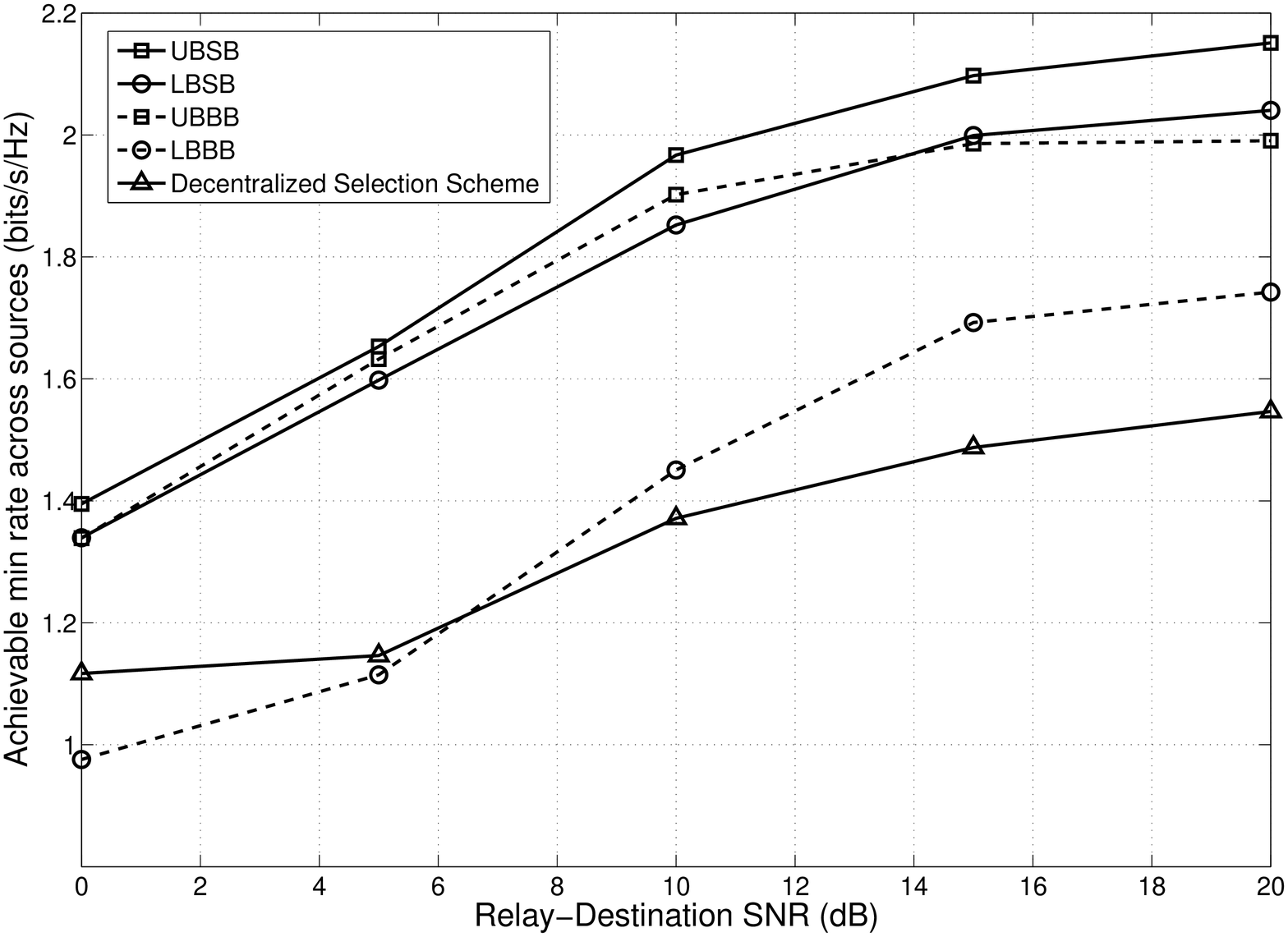}
\caption{Max-min rate across all source nodes of different resource allocation strategies in
the ``i.i.d. channel" scenario with $K = 3$, $J=2$, and $N = 8$.}
\label{fig:IID_Comp}
\end{figure}

\begin{figure}[!t]
\centering
\includegraphics[width=11cm,height=8cm]{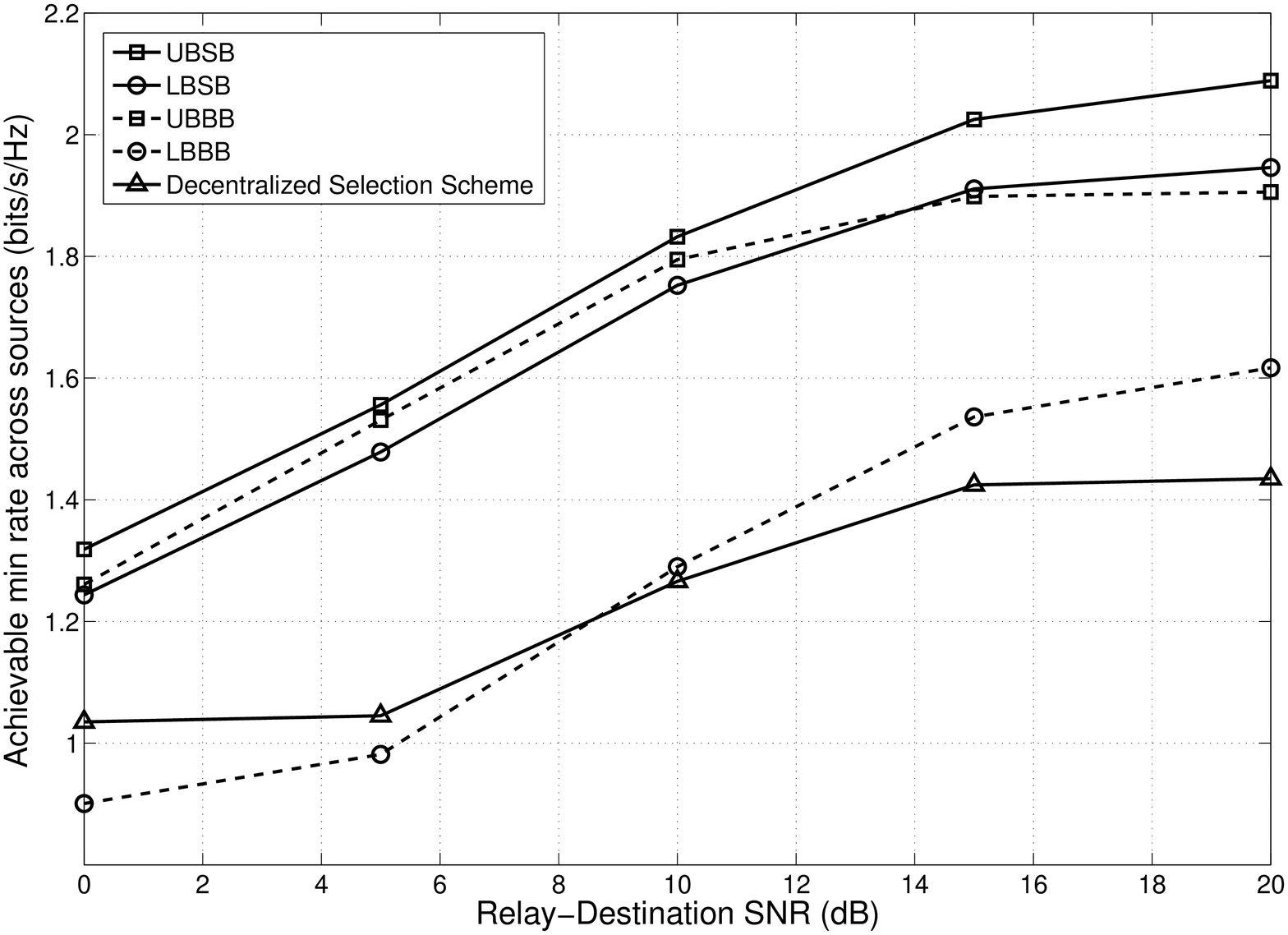}
\caption{Max-min rate across all source nodes of different resource allocation strategies in the
``i.i.d. channel" scenario with $K = 4$, $J=2$, and $N = 8$.}
\label{fig:Dist_Comp}
\end{figure}

With finite-power S-R channels, we now use the comprehensive
resource allocation and relay assignment schemes developed in
Section~\ref{sec:imperfect}. Fig.~\ref{fig:IID_Comp} ($K=3$)
and~\ref{fig:Dist_Comp} ($K=4$) plot the achievable minimum rate across
source nodes for various values of R-D SNRs. The SNR of the S-R and S-D
channels are, respectively, set to $10$dB and $5$dB. Both figures show
that at high SNRs, subcarrier-based methods outperform other resource
allocation schemes. This is expected since subcarrier-based methods
exploit the frequency diversity across relays provided by the
assumption that individual subcarriers can be transmitted
independently. However, at low SNRs, the UBBB outperforms the LBSB
scheme and the decentralized selection scheme outperforms the
centralized LBBB. This can be explained by recognizing the fact that
our heuristic method to impose selection on individual flows \textit{does not
use all available power at the relay nodes}. If we apply a second round
of power allocation at the relays, power freed up from enforcing
selection can be distributed amongst all other source nodes which are
assigned to those relays and a tighter lower bound will result.

\subsection{Resource Allocation with Finite-Power S-R Channels in Distributed
Scenario}
\begin{figure}[!t]
\centering
\includegraphics[width=11cm,height=8cm]{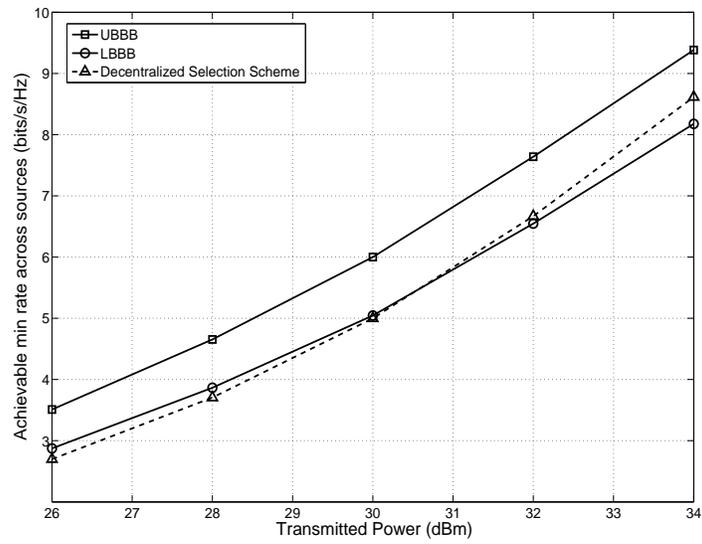}
\caption{Max-min rate across all source nodes of different resource allocation strategies in the
``distributive" scenario with $K = 3$, $J=2$, and $N = 8$.}
\label{fig:dist}
\end{figure}

Fig.~\ref{fig:dist} plots the minimum rate across users versus the
maximum available power of the sources and relays when the nodes are
geographically distributed and channels are simulated using the
COST-231 model. Although the decentralized scheme uses only local CSI,
it has a close-to-optimal performance. This method also decreases the
computation and coordination burden of the network. Again, since LBBB
does not use the total available power, it is probable that its
achievable rate is less than that of the decentralized scheme.

\section{Conclusion} \label{sec:conclusion}
This paper developed subcarrier and block-based relaying methods for the selection-based OFDM networks to maximize the
minimum rate across sources.  Considering subcarriers as independent transmissions and assuming that relays are capable of decoding all received signals, we proved that selection is violated in a maximum of $J-1$ out of $N$ subcarriers. Furthermore, by relaxing the ideal S-R channel assumption, we proposed a generalized subcarrier-based relaying scheme which jointly selects the transmission strategy, assigns relays, and allocates power to source nodes. We also characterized tight lower bounds on the minimum achievable rates of both subcarrier-based algorithms.  

However, given the implementation restrictions that IFFT/FFT pair imposed on the transmitted OFDM signals, considering individual subcarriers as independent transmissions is invalid. To hurdle this issue, we then
considered block-based relaying for multi-source networks. While heuristic algorithms with different computation complexities are proposed for the ideal S-R scenario, we formulate the optimization problem for the comprehensive case which simultaneously solves transmission strategy
selection, relay assignment, and power allocation
problem. Similar to the subcarrier-based method, tight lower bounds are also developed. 

We also proposed a simple, decentralized, relaying
algorithm and the required guidelines to allocate network-wide resources amongst source nodes. 
Compared to the previous schemes, this method significantly
decreases the required computational complexity while it has a
close-to-optimal performance.

\bibliographystyle{IEEETran}
\bibliography{MyRef}
\clearpage

\begin{table}[!t]
\renewcommand{\arraystretch}{1.3}
\centering
\caption{Parameter Values in COST-231}
\centering
\begin{tabular}{|r|c|c|l|}
\hline
Parameter & Value & Parameter & Value \\
\hline
AP Height & 15m & Frequency & 3.5 GHz\\
Building Spacing & 50m & Rooftop Height & 30m\\
Destination Height & 15m & Road Orientation & 90 deg.\\
Street Width & 12m & Noise PSD & -174 dBm\\
\hline
\end{tabular}
\label{tb:tabel1}
\end{table}

\end{document}